\documentclass{emulateapj}
\usepackage{amssymb}
\usepackage{apjfonts}
\usepackage{graphicx}
\usepackage{natbib}
\usepackage{rotating, multirow}
\usepackage{lscape}

\newcommand{\SiIVab}{Si{\sevenrm~IV}\,$\lambda\lambda$1393,1402}

\def\CIV{C\,{\sevenrm\,IV}}

\newcommand{\CIVab}{C{\sevenrm~IV}\,$\lambda\lambda$1548,1551}

\def\MgII{Mg\,{\sevenrm\,II}}
\def\MgIIwave{Mg\,{\sevenrm II}\,$\lambda$2798}
\newcommand{\MgIIa}{Mg{\sevenrm II}\,$\lambda$2796}

\newcommand{\MgIIab}{Mg{\sevenrm~II}\,$\lambda\lambda$2796,2803}

\def\zabs{$z_{\rm abs}$}
\def\zem{$z_{\rm em}$}
\def\kms{$\rm km\,s^{-1}$}
\def\ergs{${\rm erg\,s^{-1}}$}

 \font\sevenrm=cmr7 scaled 1000

\begin{document}
\title{The associated absorption features in quasar spectra of the Sload Digital Sky survey .I. \MgII\ absorption doublets}
\shorttitle{The SDSS quasar-associated absorption features}
\shortauthors{Chen et al.}

\author{Zhi-Fu Chen\altaffilmark{1}, Wei-Rong Huang\altaffilmark{1}, Ting-Ting Pang\altaffilmark{1}, Hong-Yan Huang\altaffilmark{2}, Da-Sheng Pan\altaffilmark{3}, Min Yao\altaffilmark{1}, Wei-Jing Nong\altaffilmark{1}, Mei-Mei Lu\altaffilmark{1}}

\altaffiltext{1}{Department of Physics and Telecommunication Engineering, Baise University, Baise 533000, China; zhichenfu@126.com}
\altaffiltext{2}{Department of Physics, Yunnan Normal University, Kunming 650500, China}
\altaffiltext{3}{Department of Information Technology, Guangxi Financial Vocational College, Nanning 530007, China}

\begin{abstract}
Using the SDSS spectra of quasars included in the DR7Q or DR12Q catalogs, we search for \MgIIab\ narrow absorption doublets in the spectra data around \MgIIwave\ emission lines. We obtain $17~316$ \MgII\ doublets, within the redshift range of $0.3299\le z_{\rm abs} \le 2.5663$. We find that a velocity offset of $\upsilon_r < 6000$ \kms\ is a safe boundary to constrain the vast majority of associated \MgII\ systems, although we find some doublets at $\upsilon_r > 6000$ \kms. If associated \MgII\ absorbers are defined by $\upsilon_r < 6000$ \kms, $\sim$33.3\% of the absorbers supposed to be contaminants of intervening systems. Removing the 33.3\% contaminants, $\sim$4.5\% of the quasars present at least one associated \MgII\ system with $W_{\rm r}^{\lambda2796}\ge0.2$ \AA. The fraction of associated \MgII\ systems with high velocity outflows correlates with the average luminosities of their central quasars, indicating a relationship between outflows and the quasar feedback power. The $\upsilon_r$ distribution of the outflow \MgII\ absorbers is peaked at 1023 \kms, which is smaller than the corresponding value of the outflow \CIV\ absorbers. The redshift number density evolution of absorbers ($dn/dz$) limited by $\upsilon_r > -3000$ \kms\ differs from that of absorbers constrained by $\upsilon_r > 2000$ \kms. While, absorbers limited by $\upsilon_r > 2000$ \kms\ and higher values exhibit similar profile of $dn/dz$. In addition, the $dn/dz$ is smaller when absorbers are constrained with larger $\upsilon_r$. The distributions of equivalent widths, and the ratio of $W_r^{\lambda2796}/W_r^{\lambda2803}$ is the same for associated and intervening systems, and independent on quasar luminosity.
\end{abstract}

\section{Introduction}
The optical-UV spectra of the quasar are generally characterized by a power-law shaped continuum, and often accompanied by a plenty of prominent and/or forbidden emission lines with different profiles. Continuum and emission lines are very important and advantageous for our insight into the quasar's structure, dynamics, environment, and so on. In recent years, the observations of integral field units (IFUs) have become very popular, which could obtain more detailed information of objects. While this novel technology is mainly popularized to nearby extended sources \cite[e.g.;][]{2002MNRAS.329..513D,2011ApJ...739L..47B,2011MNRAS.413..813C,2015ApJ...798....7B}, since it depends on the luminosity of each emission structure and spatial resolving power of the telescope. The IFU observation is very difficultly applied to obtain substructure information within the host galaxies of high redshift quasars (e.g., $z>1$). In addition, the low gas density of the circumgalactic medium (CGM) and intergalactic medium (IGM) determines that the emissions of the CGM and IGM are very difficultly detected with available facilities thought there are a few successful results by deep imaging and spectroscopy \cite[e.g.;][]{2014Natur.506...63C,2014ApJ...786..106M,2015ApJ...809..163A,2015Sci...348..779H,2015MNRAS.452.2388H,2016ApJS..226...25L,2017ApJ...848...78F,2017A&A...599A..28K,2017A&A...604A..23N}. Therefore, it is difficult to obtain a large sample of emission line data so that we can well constrain the characteristics of the quasar's CGM and IGM.

Quasar photons pass through foreground gaseous clouds along the quasar sightline, which are usually expected to produce absorption features in the quasar spectra. The foreground gaseous clouds have a wild range of locations, which can be located at any position between the quasar center emission regions and the observer. Therefore, one often expects that quasar absorption lines have a wild range of origins, which can be roughly divided into two categories, namely intervening and associated/intrinsic gas media. The intervening gas medium \cite[e.g.;][]{1986A&A...155L...8B,1991A&A...243..344B,2005pgqa.conf....5T,2008AJ....135..922K,2010ApJ...714.1521C} is beyond the gravitational well of the quasar system, and produces absorption features that have a significantly different redshift from the quasar system, which are usually called as intervening absorption systems. The associated/intrinsic gas medium would be located within the quasar host galaxy, galaxy halo and galaxy cluster, and produces absorption features with redshift similar to the quasar system  \cite[e.g.;][]{2004ApJ...613..129W,2007ApJS..171....1M,2008ApJ...679..239V,2012ApJ...748..131S}, which are generally called as associated absorption systems. Of course, if the associated absorption systems are related to the quasar outflow/wind with high velocity, they would host redshifts that are obviously smaller than the quasar emission redshifts \cite[e.g.;][]{2009NewAR..53..128C,2012ASPC..460...37C,2013MNRAS.434.3275C,2013ApJ...777...56C,2013MNRAS.434..163H}. Absorption signatures are very common in the quasar spectra, and their detections do not depend on the quasar emissions. In theory, absorption features would be marked on the quasar spectra as long as the quasar continuum emission passes through foreground gas medium before it reaches the observer. Therefore, the quasar associated absorption lines would be an efficient and important tool leading our insight into the nature of quasar dynamics, structures, gas distributions, environments, and so on.


The Sloan Digital Sky survey \cite[SDSS;][]{2000AJ....120.1579Y} is a very great project in astronomy community, which utilizes a dedicated wide-field 2.5 m telescope \cite[][]{2006AJ....131.2332G} located at Apache Point Observatory, New Mexico to image the universe in five broad bands \cite[ugriz;][]{1996AJ....111.1748F}. The SDSS started routine spectroscopy survey in 2000 April, and in the following 8 years (2000 April --- 2008 July), the Legacy Survey of the SDSS obtained 105 783 spectroscopically confirmed quasars \cite[e.g.;][]{2009ApJS..182..543A,2010AJ....139.2360S}. As the third stage of the SDSS (SDSS-III), the SDSS continued to collect data from 2008 July to 2014 June with update spectrographs \cite[][]{2013AJ....146...32S}, and the Baryon Oscillation Spectroscopic Survey (BOSS), which is the main dark time of the Legacy Survey of the SDSS-III, obtained $\rm 297~301$ unique quasars specra \cite[e.g.;][]{2011ApJS..193...29A,2013AJ....145...10D,2017A&A...597A..79P}. In our series of works on quasar-associated absorption systems, we will make use of the spectroscopically confirmed quasars compiled from the Legacy Survey of the first three stages of the SDSS \cite[e.g.;][]{2010AJ....139.2360S,2017A&A...597A..79P} to assemble and analyze \SiIVab, \CIVab\ and \MgIIab\ absorption doublets. Basing on the line widths of their profiles, quasar-associated absorption systems can be roughly classified into narrow absorption line systems (NALs), broad absorption line systems (BALs), and mini-BALs. The NALs generally show sharp profiles with full width at half maximum (FWHMs) less than a few hundred \kms. In this paper, we define NALs with FWHM $<1000$ \kms. As the first in a series of works on the SDSS quasar-associated absorption systems, this paper aims to look for \MgII\ NALs in the spectral data around \MgIIwave\ emission lines and statistically analyze the properties of the associated \MgII\ absorption systems.

Section \ref{sect:datasample} characterizes the quasar sample and spectral analysis. We present the properties of the \MgII\ absorption systems and discussions in Section \ref{sect:properties_discussions}. The  summary is presented in Section \ref{sect:summary}. In this paper, we adopt the $\rm \Lambda CDM$ cosmology with $\rm \Omega_M=0.3$, $\rm \Omega_\Lambda=0.7$, and $\rm H_0=70~km~s^{-1}~Mpc^{-1}$.

\section{The quasar samples and spectral analysis}
\label{sect:datasample}
\subsection{The quasar samples}
\label{sect:qsosample}
The Legacy surveys of the SDSS spectroscopically mapped objects from optical to near-infrared band at a resolution of $R\approx2000$.  The first (SDSS-I, 2000 --- 2005) and second (SDSS-II, 2005 --- 2008) stages of the SDSS produced spectra in a wavelength range of $\rm \lambda = 3800$ --- 9200 \AA\ \cite[][]{2009ApJS..182..543A}. The SDSS-I/II obtained 105 783 spectroscopically confirmed quasars \cite[DR7Q;][]{2010AJ....139.2360S}, which have absolute magnitude $M_{\rm i} < -22.0$ mag, apparent magnitude at i-band $i>15$ mag, at least one emission line with $\rm FWHM >1000$ \kms or interesting/complex absorption features, and robust emission redshifts. In the third stage of the SDSS, the BOSS produces spectra with coverage from $\lambda = 3600$ to 10400 \AA\ \cite[][]{2015ApJS..219...12A}. The BOSS had spectroscopically mapped $\rm 297~301$ unique quasars, which have $M_{\rm i} < -20.5$ Mag, at least one emission line with FWHM larger than 500 \kms or interesting/complex absorption features \cite[DR12Q;][]{2017A&A...597A..79P}. Our quasar sample is come from the quasar catalogs of DR7Q and DR12Q. In this work, for the quasars included in DR7Q, we adopt the improved redshifts of the SDSS quasars from \cite{2010MNRAS.405.2302H}. For the quasars included in DR12Q, we adopt the \MgII\ emission line based redshifts from \cite{2017A&A...597A..79P} when available, otherwise we utilize the visual inspection redshifts of \cite{2017A&A...597A..79P}. The quasar spectra are downloaded from \url{https://data.sdss.org/sas/dr12/}.

\begin{figure}
\centering
\includegraphics[width=0.41\textwidth]{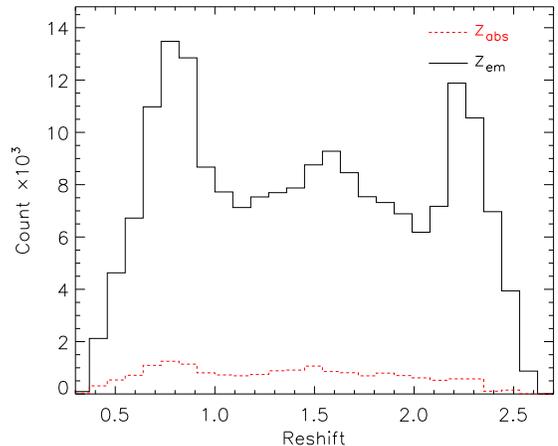}
\caption{Redshift distributions. Black solid-line represents the emission redshifts (\zem) of quasars used to search for associated \MgII\ NALs, and red dash-line is for the \MgII\ absorption redshifts (\zabs). Note that the \MgII\ absorbers are limited to the absorption systems with velocity offset $|v_{\rm r}| < 10~000$ \kms relative to the quasar systems.}
\label{fig:redshift}
\end{figure}

We aim to search for associated \MgIIab\ absorption doublets in the spectral data around \MgIIwave\ emission lines. Therefore, we firstly limit the quasar samples with $4000/2800<(1+z_{em})<8800/2800$ for the DR7Q catalog and $3800/2800<(1+z_{em})<10000/2800$ for the DR12Q catalog, where we do not consider the two end data since they are usually noisy. \cite{2017ApJ...848...79C} shows that most of the associated \MgII\ NALs are constrained within a relative velocity \footnote{$\beta \equiv \frac{\upsilon_r}{c} =\frac{(1+z_{em})^2-(1+z_{abs})^2}{(1+z_{em})^2+(1+z_{abs})^2}$, where the c is the speed of light.} $\upsilon_r<2000$ \kms. The quasar radiation has a potential to drive associated absorption clouds up to a high velocity. In order to contain vast majority of outflow absorption lines with high velocity, we search for \MgII\ narrow absorption doublets from the blue wing $\upsilon_r=10~000$ \kms\ until the red wing of \MgII\ emission line. Although there are a few outflow absorbers with $\upsilon_r>10~000$ \kms \cite[e.g.;][]{2011MNRAS.410.1957H,2013ApJ...777...56C}, we believe that the fraction of these ultra high-speed outflow absorbers would be very small. Thus, our surveyed spectral region should be safe for the search of associated \MgII\ NALs. Low signal-to-noise ratio (S/N) of the spectra often blocks the efficient survey of narrow absorption lines. Secondly, therefore we require quasar spectra with median $\rm S/N >3 ~pixel^{-1}$ in surveyed spectral region. Considering above redshift and S/N limits, we construct a large sample of $\rm 193~301$ unique quasars, which includes $\rm 198~229$ spectra, from the DR7Q and DR12Q catalogs for our survey of associated \MgII\ NALs. The quasar spectra parameters are listed in Table \ref{Tab:tableqso}, and corresponding redshifts are distributed with black solid-line in Figure \ref{fig:redshift}.

\begin{table}[htbp]
\caption{Catalog of quasar spectra searched for $\rm Mg~II$ absorption systems} \tabcolsep 0.6mm \centering 
\label{Tab:tableqso}
 \begin{tabular}{ccccccccccc}
 \hline\hline\noalign{\smallskip}
SDSS NAME & PLATE & MJD & FIBER & $\rm z_{em}$ & $L_{\rm 3000}$&median S/N\\
&&&&&\ergs& $pixel^{-1}$\\
\hline\noalign{\smallskip}
000000.97+044947.2	&	4415	&	55831	&	464	&	1.6188 	&	44.865	&	6.6 	\\
000001.27-020159.7	&	4354	&	55810	&	678	&	1.3604 	&	44.979	&	11.4 	\\
000001.37-011930.0	&	4354	&	55810	&	646	&	2.3280 	&	45.077	&	3.2 	\\
000001.93-001427.4	&	4216	&	55477	&	312	&	2.1630 	&	45.085	&	5.0 	\\
000002.15+151516.6	&	6172	&	56269	&	394	&	1.7101 	&	44.888	&	7.6 	\\
000003.28+105744.5	&	6182	&	56190	&	672	&	1.8190 	&	44.790	&	4.4 	\\
000003.94+263645.6	&	6877	&	56544	&	564	&	2.1801 	&	45.476	&	13.0 	\\
000005.13+083740.1	&	6152	&	56164	&	23	&	2.0014 	&	45.289	&	9.4 	\\
000006.01-035334.1	&	7034	&	56564	&	636	&	0.6754 	&	44.023	&	5.5 	\\
000006.42+324335.3	&	7144	&	56564	&	266	&	2.2100 	&	44.898	&	3.9 	\\
\hline\hline\noalign{\smallskip}
\end{tabular}
\\
\end{table}

\subsection{The spectral analysis}
\label{sect:spectral_analysis}
We adopt consistent methods used in our previous works \cite[e.g.;][]{2014ApJS..210....7C,2014ApJS..215...12C,2015ApJS..221...32C} to search for \MgII\ absorption doublets, which contains several steps: 1) the pseudo-continuum fits; 2) the surveys of absorption candidates; 3) the measurements of absorption line parameters. We briefly describe these primary processes in the follows.
\begin{enumerate}
  \item The pseudo-continuum fits. A combination of cubic spline and multi-Gaussian functions is invoked to model the underlying continuum plus emission lines, which is called as pseudo-continuum, in an iterative fashion \cite[e.g.;][]{2005ApJ...628..637N,2011AJ....141..137Q}. The absorption candidates are searched in the spectral data normalized by the pseudo-continuum fit. The continual absorption features with widths larger than 2000 \kms\ at depths larger than 10\% under the pseudo-continuum fit, which are generally called as BALs \cite[e.g.;][]{1979ApJ...234...33W}, are marked and disregarded by the surveyed program of narrow absorption lines. As an example, Figure \ref{fig:sample} exhibits the spectrum of the quasar SDSS J145802.70+525240.8 over plotted with fitting results.

      \begin{figure}
      \centering
      \includegraphics[width=0.45\textwidth]{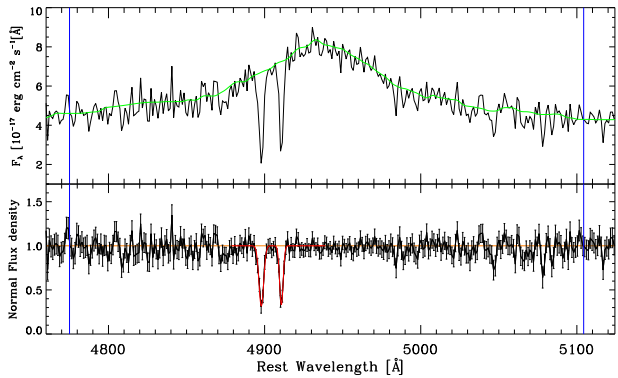}
      \caption{The spectrum of the quasar SDSS J145802.70+525240.8 with $z_{\rm em}=0.7655$. Blue vertical lines label the boundaries of surveyed spectral region of \MgII\ NALs. Upper panel: the green solid-line is the pseudo-continuum fit. Bottom panel: the quasar flux and flux uncertainty ($1\sigma$) have been normalized by the pseudo-continuum fit (black line with error bar). Red Gaussian curves indicate a \MgII\ absorption doublet at $z_{\rm abs}=0.7515$.}
      \label{fig:sample}
      \end{figure}

  \item The surveys of narrow \MgII\ absorption candidates. Absorption candidates are searched in the normalized spectra data. This process is principally controlled by the separation of the \MgIIab\ doublet.  We neglect the continual absorption features with line width between 1000 \kms and 2000 \kms\ at depths larger than 10\% under the pseudo-continuum fit and no clear two-trough profile, which are usually not well characterized with a pair of Gaussian functions and are generally belonged to mini-BALs. A pair of Gaussian functions is invoked to model each \MgII\ doublet candidate, and the fitting results are visually checked one by one. The red curve in Figure \ref{fig:sample} exhibits the fitting result of a \MgII\ absorption doublet at $z_{\rm abs}=0.7515$.

  \item The measurements of absorption line parameters. The \MgII\ absorption system redshift is determined by the Gaussian function fitting center of the \MgIIa\ line. We directly integrate the Gaussian function fitting profile to yield the equivalent width ($W$) of absorption line. One $1\sigma$ error of the equivalent width is computed via
      \begin{equation}
      \label{eq:werr}
      \sigma_w=\frac{\sqrt{\sum\limits_{i=1}^{N}
      P^2(\lambda_i-\lambda_0)\sigma^2_{f_i}}}{\sum\limits_{i=1}^{N}
      P^2(\lambda_i-\lambda_0)}\Delta\lambda,
      \end{equation}
      where $P(\lambda_i-\lambda_0)$ is the Gaussian function fitting profile centered at $\lambda_0$, $\lambda_i$ is the wavelength, $\sigma_{f_i}$ is the normalized flux uncertainty, and $\rm N$ is the pixel number over $\rm \pm3\sigma$, here $\rm \sigma$ is given by the Gaussian function fit of the given absorption line.
\end{enumerate}

We only retain the \MgII\ absorption systems with $W^{\lambda2796}\ge 3\sigma_{W^{\lambda2796}}$, $W^{\lambda2803}\ge 2\sigma_{W^{\lambda2803}}$, $W_r^{\lambda2796}\ge 0.2$ \AA, and $W_r^{\lambda2803}\ge 0.2$ \AA, where $W_r$ is the equivalent width at absorber rest frame.  Finally, we detect $\rm17 316$ reliable \MgII\ absorption doublets, which are located within $0.3299\le z_{\rm abs} \le 2.5663$. We distribute these absorption redshifts with red dash-line in Figure \ref{fig:redshift}, and tabulate the absorption line parameters in Table \ref{Tab:tableabs}.

\begin{table*}[htbp]
\caption{Catalog of $\rm Mg~II$ absorption systems} \tabcolsep 1.2mm \centering 
\label{Tab:tableabs}
 \begin{tabular}{ccccccccccc}
 \hline\hline\noalign{\smallskip}
SDSS NAME & PLATE & MJD & FIBER & $\rm z_{em}$ & $\rm z_{abs}$ & $\rm W_r^{\lambda2796}$ & $\rm W_r^{\lambda2803}$ & $v_{\rm r}$\\
&&&&&&\AA&\AA& \kms\\
\hline\noalign{\smallskip}
000002.15+151516.6	&	6172	&	56269	&	394	&	1.7101 	&	1.7190 	&	1.65 	$\pm$	0.35 	&	1.41 	$\pm$	0.21 	&	-983	\\	
000016.49+022715.1	&	4296	&	55499	&	642	&	0.8850 	&	0.8703 	&	0.97 	$\pm$	0.18 	&	0.59 	$\pm$	0.12 	&	2348	\\	
000023.25+192732.2	&	6127	&	56274	&	38	&	0.8402 	&	0.8340 	&	3.61 	$\pm$	0.54 	&	2.96 	$\pm$	0.41 	&	1012	\\	
000026.71-050334.7	&	7034	&	56564	&	346	&	0.8767 	&	0.9384 	&	0.28 	$\pm$	0.09 	&	0.26 	$\pm$	0.08 	&	-9701	\\	
000036.07+191511.5	&	6127	&	56274	&	32	&	2.2346 	&	2.2393 	&	0.77 	$\pm$	0.21 	&	0.51 	$\pm$	0.16 	&	-435	\\	
000038.79+090226.2	&	4534	&	55863	&	512	&	1.5650 	&	1.5562 	&	2.83 	$\pm$	0.39 	&	2.45 	$\pm$	0.45 	&	1031	\\	
000045.77+255106.1	&	2822	&	54389	&	339	&	1.4446 	&	1.4224 	&	0.57 	$\pm$	0.09 	&	0.51 	$\pm$	0.09 	&	2736	\\	
000057.79+294236.5	&	7134	&	56566	&	394	&	0.7597 	&	0.7689 	&	0.20 	$\pm$	0.06 	&	0.27 	$\pm$	0.13 	&	-1564	\\	
000058.22-004646.5	&	387 	&	51791	&	93	&	1.8929 	&	1.8861 	&	0.98 	$\pm$	0.14 	&	0.67 	$\pm$	0.16 	&	706	\\	
000105.17+081907.4	&	4534	&	55863	&	464	&	0.8210 	&	0.8204 	&	4.04 	$\pm$	0.50 	&	3.09 	$\pm$	0.45 	&	98	\\	
\hline\hline\noalign{\smallskip}
\end{tabular}
\\
\end{table*}

\subsection{Missed systems}
\label{Sect:missed_systems}
Our program searched for \MgII\ candidates is governed by the separation of the \MgII\ doublet, and each candidate is visually checked one by one. During the manually checked process, we reject the the potential absorptions that are blended with significant noise, without obvious absorption profile, or can not be well modeled by a pair of Gaussian functions. This manually checked process is a very hard work, and depended on personal experience of dealing with absorption lines. Therefore, it is inevitable that there are a few missed systems made by human errors, though we believe this portion would be small. In order to quantify our incidence of missed \MgII\ systems, we firstly check the associated \MgII\ systems included in \cite{2008ApJ...679..239V} and \cite{2012ApJ...748..131S}, and find that $>96\%$ and $>99\%$ of their associated \MgII\ systems are recovered by our campaigns. Most of the missed systems are weak absorptions or fallen into noise spectral regions. Secondly, two members of our group independently inspect 2000 quasar spectra to search for accepted \MgII\ doublets. In the entire sample of 2000 quasar spectra, we find that there are 3 missed systems. In term of the comparisons with previous works and our internal cross-check, we conclude that the fraction of missed systems caused by human error would be very small. The systematic human error would not be correlated with \MgII\ system properties.

\section{\MgII\ absorber properties and discussions}
\label{sect:properties_discussions}
\subsection{Velocity distribution of absorbers}
\label{sect:dndv}
The quasar is located within its host galaxy, which is surrounded by CGM and would be a member galaxy of a galaxy cluster. The quasar center emission likely passes through its surrounding gas clouds located within outflow, host galaxy, CGM and IGM of galaxy cluster before it reaches the observer, and produces marks in the quasar spectra in a form of absorption features. Therefore, one often expects that the quasar absorption systems would exhibit a clustering distribution around the quasar emission redshift relative to the cosmologically absorption systems. This phenomenon has been confirmed by previous works \cite[e.g.,][]{2008MNRAS.386.2055N,2008MNRAS.388..227W,2016MNRAS.462.3285P,2015ApJS..221...32C,2017ApJ...848...79C}. In this section, we will statistically analyze the distributions of the absorber number density ($dn/d\upsilon$) around quasars, where the $dn/d\upsilon$ is defined as the number of absorbers per unit of velocity interval, and investigate the dependence of the $dn/d\upsilon$ on quasar luminosity.

\subsubsection{The natures of absorbers}
\label{sect:dndv_nature}
\begin{figure}
\centering
\includegraphics[width=0.45\textwidth]{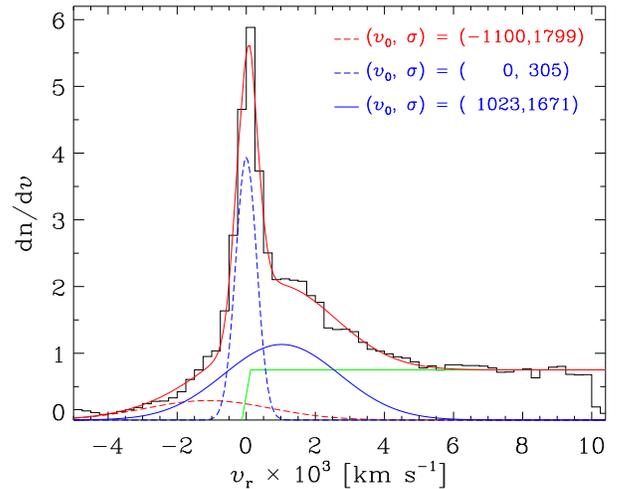}
\caption{Relative velocity distributions of absorbers. Red dash-, blue dash- and blue solid-curves indicate the Gaussian function fits, which correspond to inflow, environment and outflow absorbers, respectively. Green solid-line indicates the mean count at $\upsilon_r>6000$ \kms, which corresponds to intervening absorbers. Red solid-curve is the sum of the all color lines. The values shown in the top-right corner are the Gaussian function fitting centers ($\upsilon_0$) and dispersions ($\sigma$).}
\label{fig:dopp_dist}
\end{figure}

This paper searched for associated \MgII\ absorption systems in quasar spectra data with $\upsilon_r \le 10~000$ \kms. \cite{2017ApJ...848...79C} has claimed that most of the associated \MgII\ absorbers are constrained within $\upsilon_r < 2000$ \kms, though some outflow absorbers with high velocity \cite[e.g.;][]{2013ApJ...777...56C,2013MNRAS.434..163H} would be beyond this limit. Therefore, the large velocity range of $\upsilon_r \le 10~000$ \kms\ guarantees that our \MgII\ absorber sample contains not only environmental absorbers but also most of the outflow/wind absorbers with high velocity, though it likely contains a substantial absorbers associated with cosmologically intervening galaxies. In Figure \ref{fig:dopp_dist}, we show the distribution of the number density of the absorbers included in Table \ref{Tab:tableabs}. It is clear that there is an obvious excess around $\upsilon\approx0$, which implies that the \MgII\ absorbers are clustered around quasars. In addition, the complex distribution of the $dn/d\upsilon$ clearly implies that our \MgII\ absorber sample should assuredly include absorption systems originated in quasar surrounding environment, outflow/wind, and foreground intervening galaxies.

Figure \ref{fig:dopp_dist} obviously shows a uniform random distribution at $\upsilon_r > 6000$ \kms, which would be dominated by intervening absorbers. We adopt the mean count of absorbers with $\upsilon_r > 6000$ \kms\ to account for the $dn/d\upsilon$ distribution of the intervening absorbers, and extend it to $\upsilon_r = 0$ (\textbf{see the green line of Figure \ref{fig:dopp_dist}}). The quasar environmental absorbers should exhibit a normal $dn/d\upsilon$ distribution located at $\upsilon_r = 0$, and the outflow/wind absorbers likely destroy this normal distribution and arise a bump at blue wing ($\upsilon_r > 0$). Figure \ref{fig:dopp_dist} exhibits an extended red wing ($\upsilon_r < 0$), which is possibly contributed from inflow absorbers. Here we adopt three Gaussian functions to characterize the $dn/d\upsilon$ distributions of the inflow (Gaussian center $\upsilon_r < 0$), environment (Gaussian center $\upsilon_r = 0$) and outflow/wind (Gaussian center $\upsilon_r > 0$) absorbers, and adopt a linear function with slope $\alpha = 0$ and intercept equating to the mean count of absorbers with $\upsilon_r > 6000$ \kms\ to account for the $dn/d\upsilon$ distribution of the intervening absorbers. The fitting results are shown in Figure \ref{fig:dopp_dist}, where the red dash-, blue dash-, blue solid-, and green solid-lines correspond to the inflow, environment, outflow/wind, and intervening absorbers, respectively. We find that the population of the outflow/wind absorbers has a peak at $\upsilon_r = 1023$ \kms\ and is extended beyond $\upsilon_r > 6000$ \kms. The velocity of the peak position of the \MgII\ outflow/wind absorbers is significantly smaller than 2000 \kms, which is the peak position of the \CIV\ outflow/wind absorbers \cite[e.g.,][]{2015MNRAS.450.3904C}. This suggests that the \CIV\ outflow/wind absorbers have a higher velocity relative to the \MgII\ ones. These different velocities of the outflow/wind absorbers likely relate to the ionization potentials and ionization mechanisms of the \CIV\ and \MgII\ ions. In the scheme of ionization potentials, the \CIV\ absorbers with high ionization potential would be likely closer to the quasar center region with respect to the \MgII\ absorbers with low ionization potential. Therefore with respect to the \MgII\ absorbers, the \CIV\ absorbers would be accelerated by stronger quasar radiation pressure and have a higher velocity. In the scheme of ionization mechanisms, the ionizations of the \MgII\ gas might be dominated by photoionization. While, in addition to the photoionization, a significant portion of the \CIV\ gas might be ionized by shocks, which can accelerate the \CIV\ absorbers to a higher velocity as well. The detail mechanisms, which result in different velocity offsets for the \MgII\ and \CIV\ outflow/wind absorbers, are beyond the topic of this paper.

\begin{figure}
\centering
\includegraphics[width=0.45\textwidth]{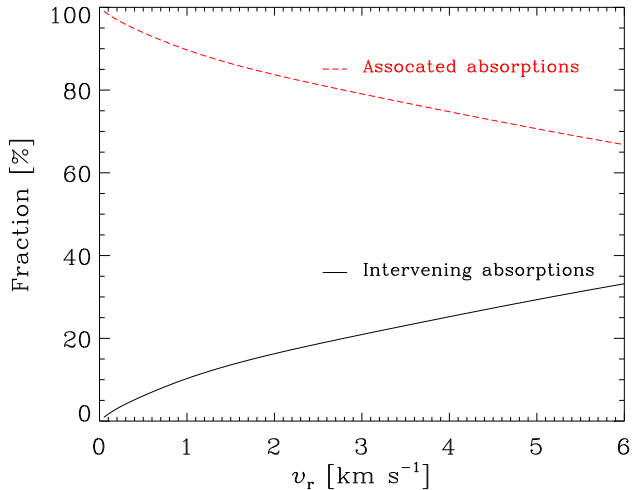}
\caption{The ratios of the numbers of the purely associated (red dash-curve) or intervening (black solid-curve) absorbers to the total number of absorbers included in the associated absorber sample, as a function of boundary velocity used to defined the associated absorber sample.}
\label{fig:ass_int_fraction}
\end{figure}

Figure \ref{fig:dopp_dist} suggests that the \MgII\ outflow/wind absorbers are limited within $\upsilon_r < 6000$ \kms, though there would be a few events with a higher velocity \cite[e.g.;][]{2013ApJ...777...56C,2013MNRAS.434..163H}. Thus, the $\upsilon_r = 6000$ \kms\ would be a very safe boundary to divide the quasar associated and intervening \MgII\ absorbers. In other words, a vast majority of associated \MgII\ absorbers should have a velocity $\upsilon_r < 6000$ \kms. While, we also note from Figure \ref{fig:dopp_dist} that a sample of associated \MgII\ absorbers defined by the $\upsilon_r < 6000$ \kms\ would be significantly contaminated by intervening absorbers. Here we roughly estimate the fractions of the inflow, environment, outflow and intervening absorbers within $\upsilon_r < 6000$ \kms, by calculating the ratios of the areas encircled by the color curves shown in Figure \ref{fig:dopp_dist}. We find that within $\upsilon_r < 6000$ \kms, the fractions of the inflow (area encircled by the red dash-curve), environment (area encircled by the blue dash-curve), outflow (area encircled by the blue solid-curve) and intervening (area encircled by the green solid-line) absorbers are 9.7\%, 22.1\%, 34.9\% and 33.3\%, respectively. We define the purely associated absorptions as the total absorptions of the inflow, environment and outflow. We note that the fraction of the purely associated component is only 66.7\% (9.7\% + 22.1\% + 34.9\%), and the fraction of the intervening component is large (33.3\%). In other words, the associated absorber sample, if it is defined by the $\upsilon_r < 6000$ \kms, is significantly contaminated by the intervening absorbers. Figure \ref{fig:ass_int_fraction} shows the fractions of the purely associated (red-dash curve) and intervening (black solid-curve) \MgII\ absorbers as a function of boundary used to define the associated absorber sample.

\subsubsection{Dependence on UV continuum luminosity}
\label{sect:dndv_L}
\begin{figure*}
\centering
\includegraphics[width=0.45\textwidth]{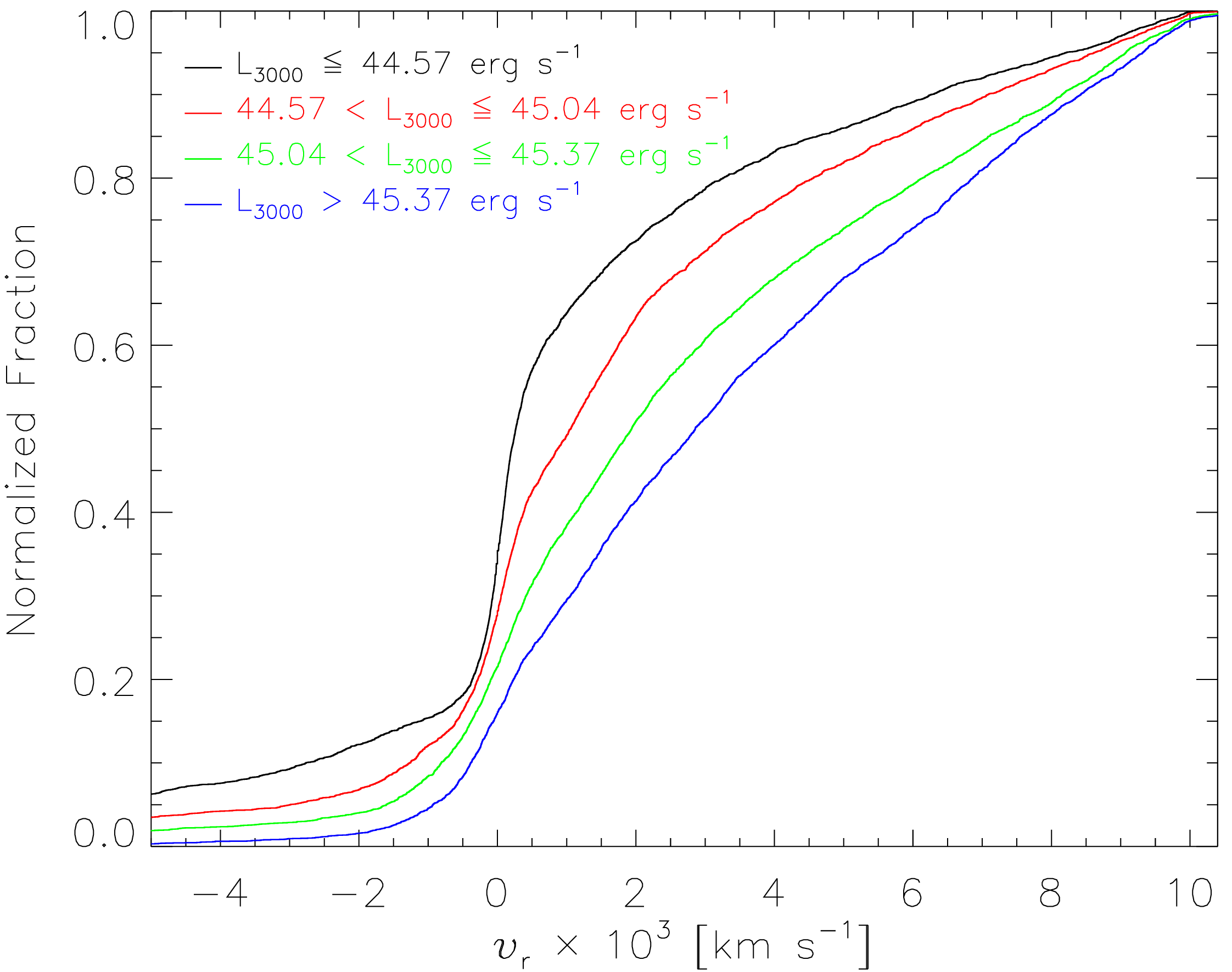}
\hspace{2ex}
\includegraphics[width=0.45\textwidth]{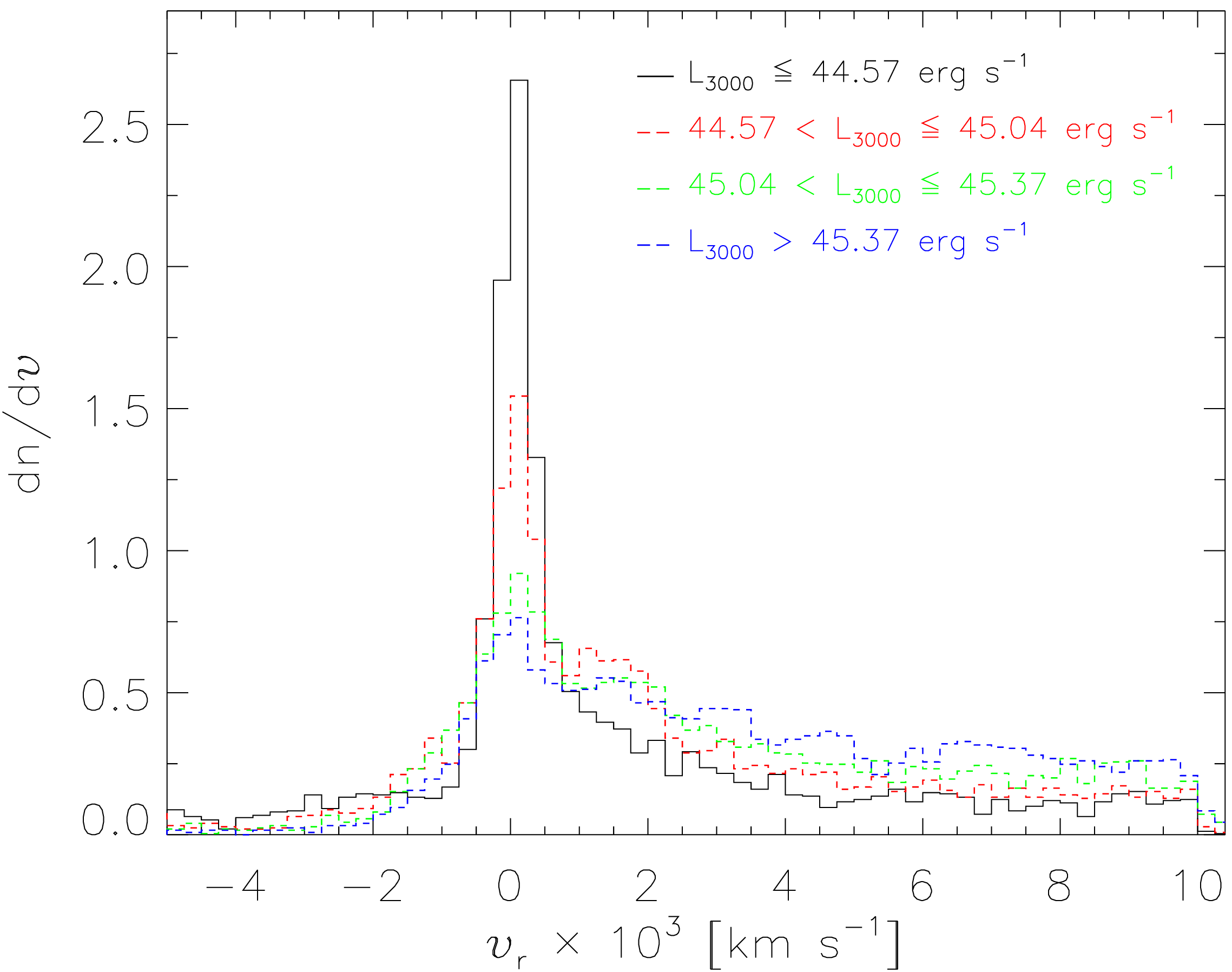}
\caption{Relative velocity distributions of absorbers. Left panel: y-axis is the cumulative counts of absorbers, which have been normalized by the total numbers of absorbers within each subsample. Different color curves represent quasars with different luminosities at rest frame 3000 \AA. Right panel is consistent with middle panel but shown with histograms.}
\label{fig:dopp_L}
\end{figure*}

Many previous researches \cite[e.g.,][]{2009ApJ...697..345C,2011A&A...525A..51B,2012ApJ...754...38C,2013MNRAS.435..346J} have claimed that the spectra of different kind of background sources, such as blazars, Gamma-ray busts (GRBs) and normal quasars, would show inconsistent incident rates of intervening absorptions. This is likely connected to the center radiations of the background sources, which would play an important role to the outflows. The strong radiation sources are expected to host outflow with high velocity, and thus have a high probability polluting the incident rate of intervening absorptions. Here, basing on the quasar luminosities at rest frame 3000 \AA\ ($L_{3000}$), we divided our \MgII\ absorber sample into four subsamples with similar absorber numbers and check the absorber velocity distributions. The results are shown in Figure \ref{fig:dopp_L}. It is clear that the absorber velocity distributions are obviously related with the quasar luminosities. The quasars with the higher luminosities are expected to exhibit the higher fraction of outflow absorbers and the longer tail of extended velocity. For example, the outflow absorbers of the quasars with $L_{3000} \le 44.57$ \ergs\ are limited within 4000 \kms\ (black solid-line in the right panel of Figure \ref{fig:dopp_L}), while those with $L_{3000} > 45.37$ \ergs\ are well extended beyond 6000 \kms\ (blue dash-line in the right panel of Figure \ref{fig:dopp_L}). This implies that the quasar outflow is positively correlated with the quasar feedback power.

\subsection{Coverage fraction and redshift number density of absorbers}
\label{sect:CF_dndz}
In this paper, we have constructed a large sample of $\rm 193~301$ quasars to search for \MgII\ absorption doublets with $\upsilon_r \le 10~000$ \kms.
We find that among the $\rm 193~301$ quasars, there are $\rm 15~925$ quasars with at least one \MgII\ absorption systems. In Section \ref{sect:dndv}, we have found that the vast majority of associated \MgII\ absorbers would be constrained within $\upsilon_r < 6000$ \kms. Using this $\upsilon_r < 6000$ \kms\ boundary to limit our absorber catalog, we find that $\rm 13~163$ quasars are observed at least one \MgII\ absorption systems, which suggests that the fraction of quasars with \MgII\ absorption systems with $\upsilon_r < 6000$ \kms is about 6.8\%. Section \ref{sect:dndv} has also claimed that within $\upsilon_r < 6000$ \kms, about 66.7\% of the \MgII\ absorption systems are possibly formed in quasar's inflows, outflows/winds and environments. In other words, about 33.3\% of the \MgII\ absorbers would be located within foreground cosmologically intervening galaxies. Accounting for this fraction of the purely associated absorbers, we find that only $6.8\%\times66.7\%\approx4.5\%$ of the quasars are observed at least one associated \MgII\ absorption systems.

\begin{figure}
\centering
\includegraphics[width=0.45\textwidth]{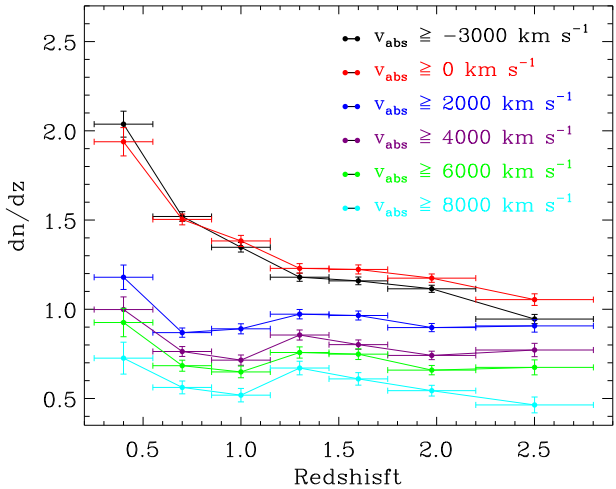}
\caption{Redshift number density evolution of the \MgII\ absorbers. The vertical error bars are from the Poisson statistics. The different color symbols represent the absorbers located within different relative velocity range from the quasars.}
\label{fig:dndz}
\end{figure}

Figure \ref{fig:dopp_L} clearly shows that the associated absorbers of the faint quasars are apt to be concentrated at a low relative velocity, while many absorbers of the luminous quasars host high velocities. This would be attributed to the strong radiations of the luminous quasars, which accelerate the absorbers to high velocities. The associated absorbers with high velocity can pollute the cosmologically intervening population, so that the polluted region likely exhibits a higher incident rate of absorbers when compared to the region with very large $\upsilon_r$.
Here we estimate the incident rate of \MgII\ absorbers ($dn/dz$) in different velocity range to check the velocity extension of the associated absorbers. The $dn/dz = N_{\rm abs}/\Delta z$, where the $N_{\rm abs}$ is the total number of absorbers with $W_r^{\lambda2796}\ge 0.2$ \AA\ and detected within the surveyed redshift path $\Delta z$:
\begin{equation}
\label{eq:zpath}
 \Delta z(W_r^{\lambda}) = \int_{z_{min}}^{z_{max}} \sum_i^{N_{spec}} g_i(W_r^{\lambda},z)dz,
\end{equation}
where $z_{min}$ and $z_{max}$ are jointly determined by the given redshift range, surveyed spectra region and the limits of spectra data, $g_i(W_r^{\lambda},z)=1$ if the $W_r^{\lambda}$ is larger than the detection threshold, otherwise $g_i(W_r^{\lambda},z)=0$, and the sum is over all quasars. The error of the $dn/dz$ can be determined from Poisson statistics. The results are displayed in Figure \ref{fig:dndz}.

Figure \ref{fig:dndz} clearly exhibits some interesting information. Firstly, the $dn/dz$ evolution profiles are similar for the absorbers with $\upsilon_r > 2000$ \kms, but differ from those that also account for the absorbers with lower velocity offsets. \cite{2017ApJ...848...79C} has claimed that most of the \MgII\ associated absorbers are constrained within $\upsilon_r < 2000$ \kms, which is also supported by Figure \ref{fig:dopp_dist}. In other words, the population of absorbers with $\upsilon_r > 2000$ \kms\ is dominated by intervening absorbers, though there are some outflow absorbers with $\upsilon_r > 2000$ \kms. This would be an important reason why the absorbers with $\upsilon_r > 2000$ \kms\ have similar $dn/dz$ evolution profile. In addition, it also explains that the significant contribution from associated absorbers results in a $dn/dz$ evolution profile of absorbers with $\upsilon_r > -3000$ \kms\ or $\upsilon_r > 0$ \kms, which differs from that of absorbers with $\upsilon_r > 2000$ \kms. Secondly, for the absorbers with $\upsilon_r > 2000$ \kms, the $dn/dz$ of absorbers limited by smaller $\upsilon_r$ is bigger than that of absorbers constrained by larger $\upsilon_r$. This can be ascribed to the associated absorbers as well. The contribution from associated absorbers could not change the $dn/dz$ evolution profiles of absorbers with $\upsilon_r > 2000$ \kms, while it can result in bigger $dn/dz$. The larger velocity limit the smaller fraction of associated absorbers is, and the smaller $dn/dz$ is.

\subsection{Equivalent width distributions}
\label{sect:ew}
\begin{figure*}
\centering
\includegraphics[width=0.31\textwidth]{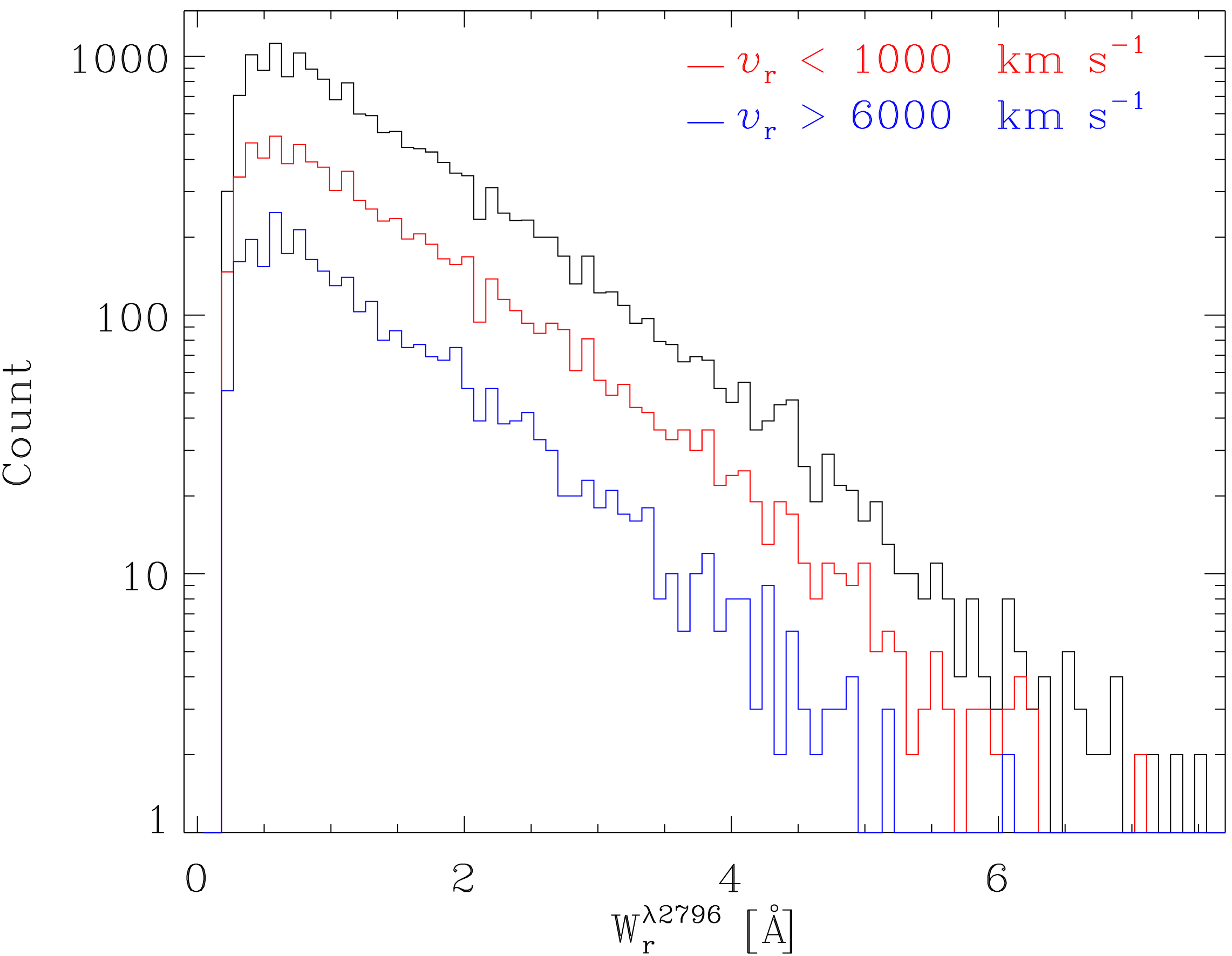}
\hspace{2ex}
\includegraphics[width=0.31\textwidth]{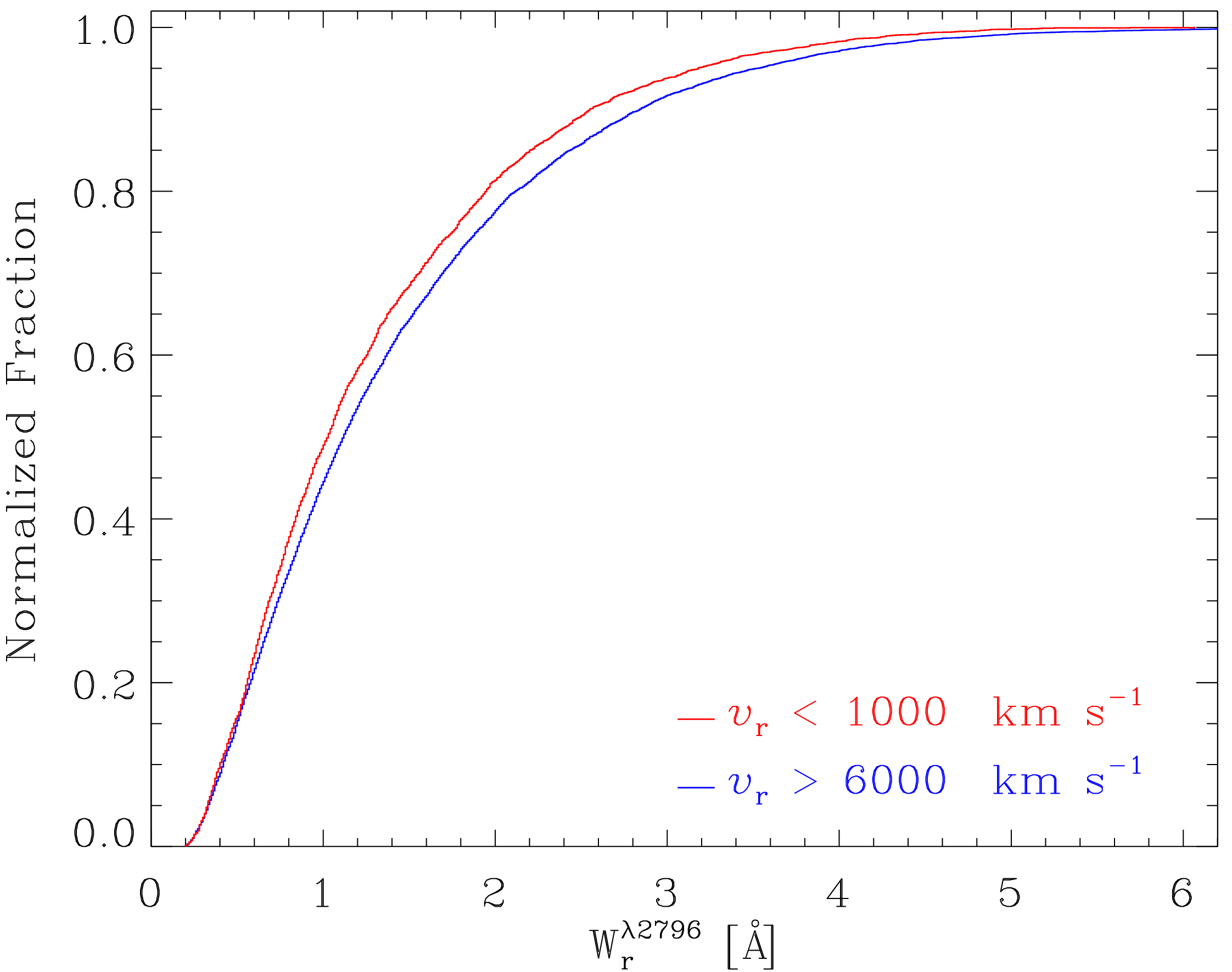}
\hspace{2ex}
\includegraphics[width=0.31\textwidth]{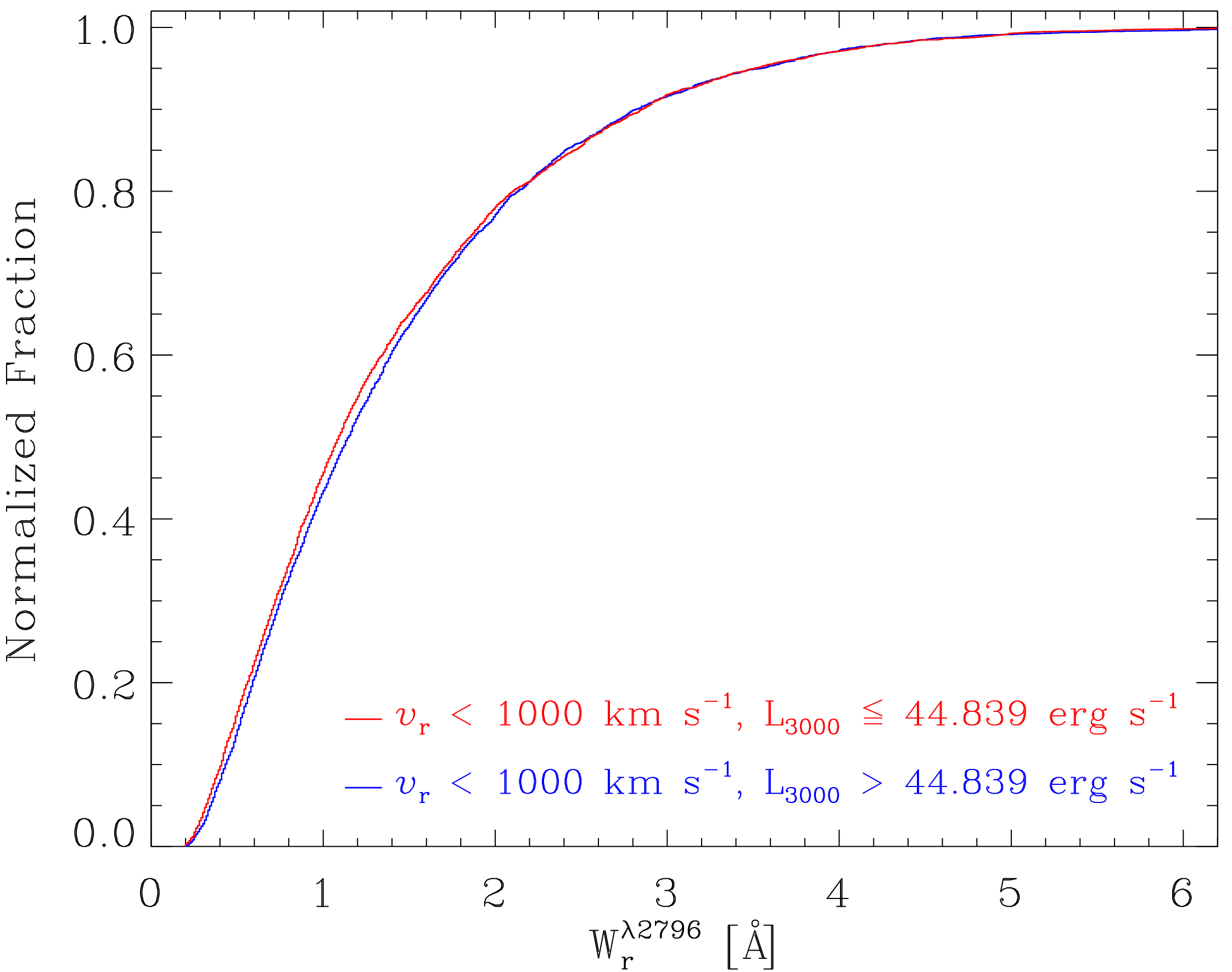}
\caption{Distributions of the absorption strengths at rest frame. Left panel: black line is for the all absorbers included in our absorber catalog, red line is for the absorbers with $\upsilon_r < 1000$ \kms, and blue line is for the ones with $\upsilon_r > 6000$ \kms. Middle panel: same as the color lines shown in left panel. The y-axis is normalized by the total number of absorbers included in each subsample. Right panel: all absorbers are limited with $\upsilon_r < 1000$ \kms. Red line is for the absorbers of the quasars with luminosities less than the median value of the $L_{\rm 3000}$, and blue line is for the ones of the quasars with luminosities larger than the median value of the $L_{\rm 3000}$.}
\label{fig:distr_w}
\end{figure*}

Our large absorber sample contains $\rm 17~316$ \MgII\ absorption systems with $0.3299\le z_{\rm abs} \le 2.5663$, whose absorption strengths are displayed in Figure \ref{fig:distr_w}. We find that about 56.1\% and 22.7\% of absorbers have $W_r^{\lambda2796}\ge 1$ \AA\ and $W_r^{\lambda2796}\ge 2$ \AA, respectively. Figure \ref{fig:dopp_dist} suggests that the absorbers with $\upsilon_r > 6000$ \kms\ should be dominated by cosmologically intervening absorbers. In addition, we note that although the absorbers with $\upsilon_r < 6000$ \kms\ are dominated by associated absorbers, the fraction of intervening absorbers is significant. In order to reduce the contamination of the intervening absorbers to the associated ones and compare the properties of associated and intervening absorbers, here we define the associated \MgII\ absorbers with $\upsilon_r < 1000$ \kms, where $\upsilon_r = 1000$ \kms\ is similar to the peak position of the outflow absorbers (see blue solid-line of Figure \ref{fig:dopp_dist}). The left and middle panels of Figure \ref{fig:distr_w} suggest that both associated and intervening absorbers have similar strength distributions. In addition, the right panel of Figure \ref{fig:distr_w} suggests that the absorption strengths of the associated absorbers do not obviously related with the quasar luminosities. In a word, we do not find a significant difference of absorption strengths between associated and intervening absorbers, and between faint and luminous quasars.

\begin{figure*}
\centering
\includegraphics[width=0.31\textwidth]{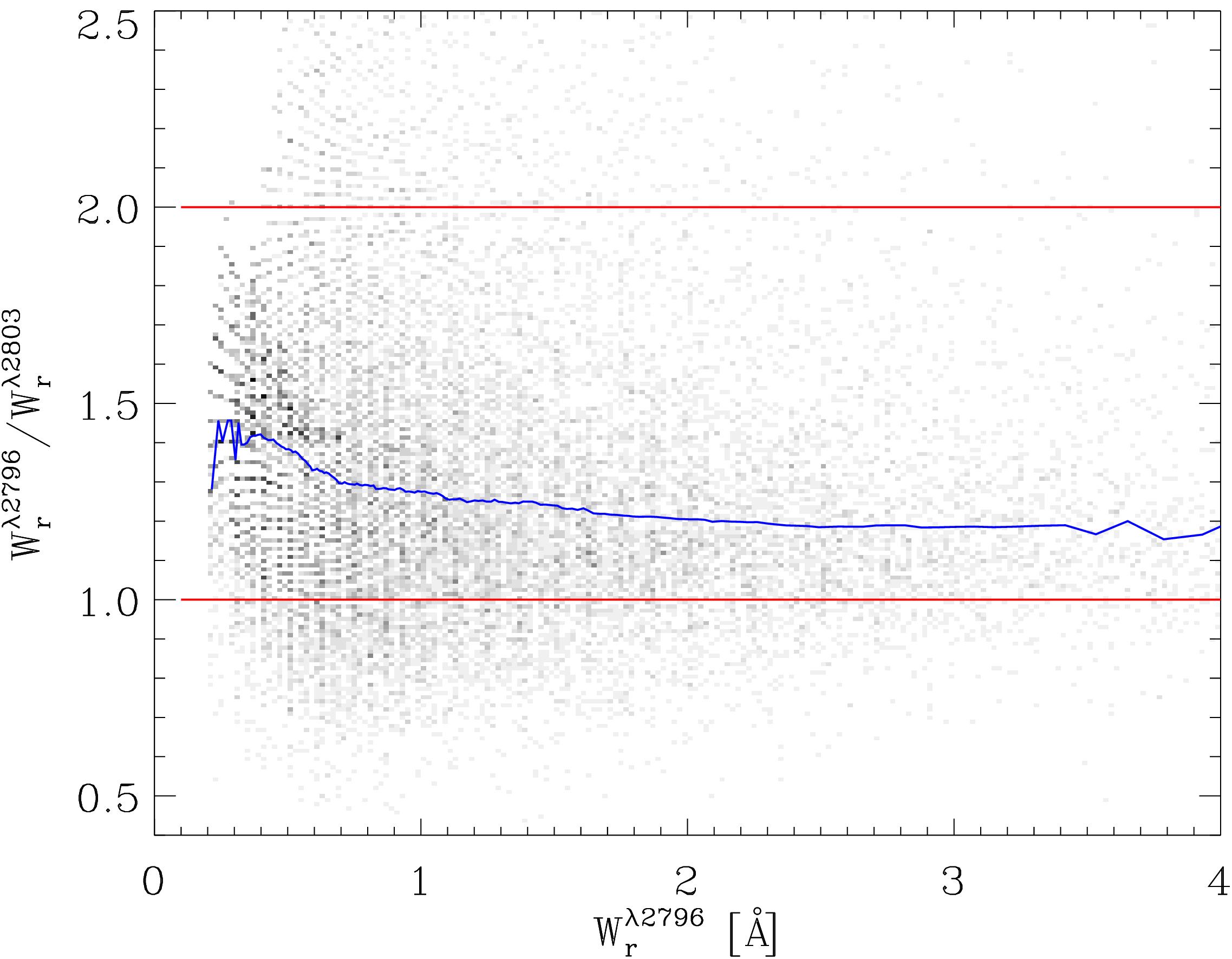}
\hspace{2ex}
\includegraphics[width=0.31\textwidth]{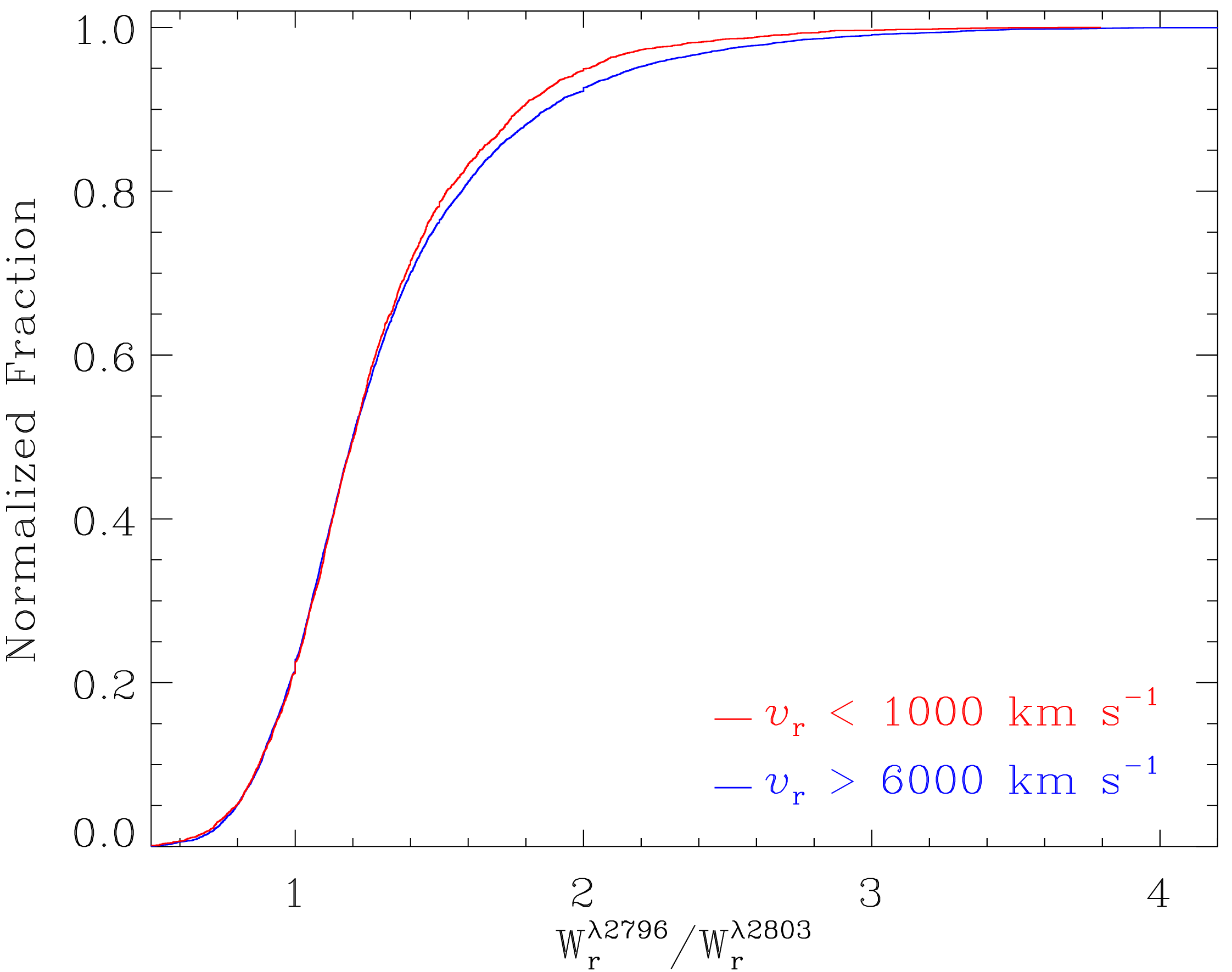}
\hspace{2ex}
\includegraphics[width=0.31\textwidth]{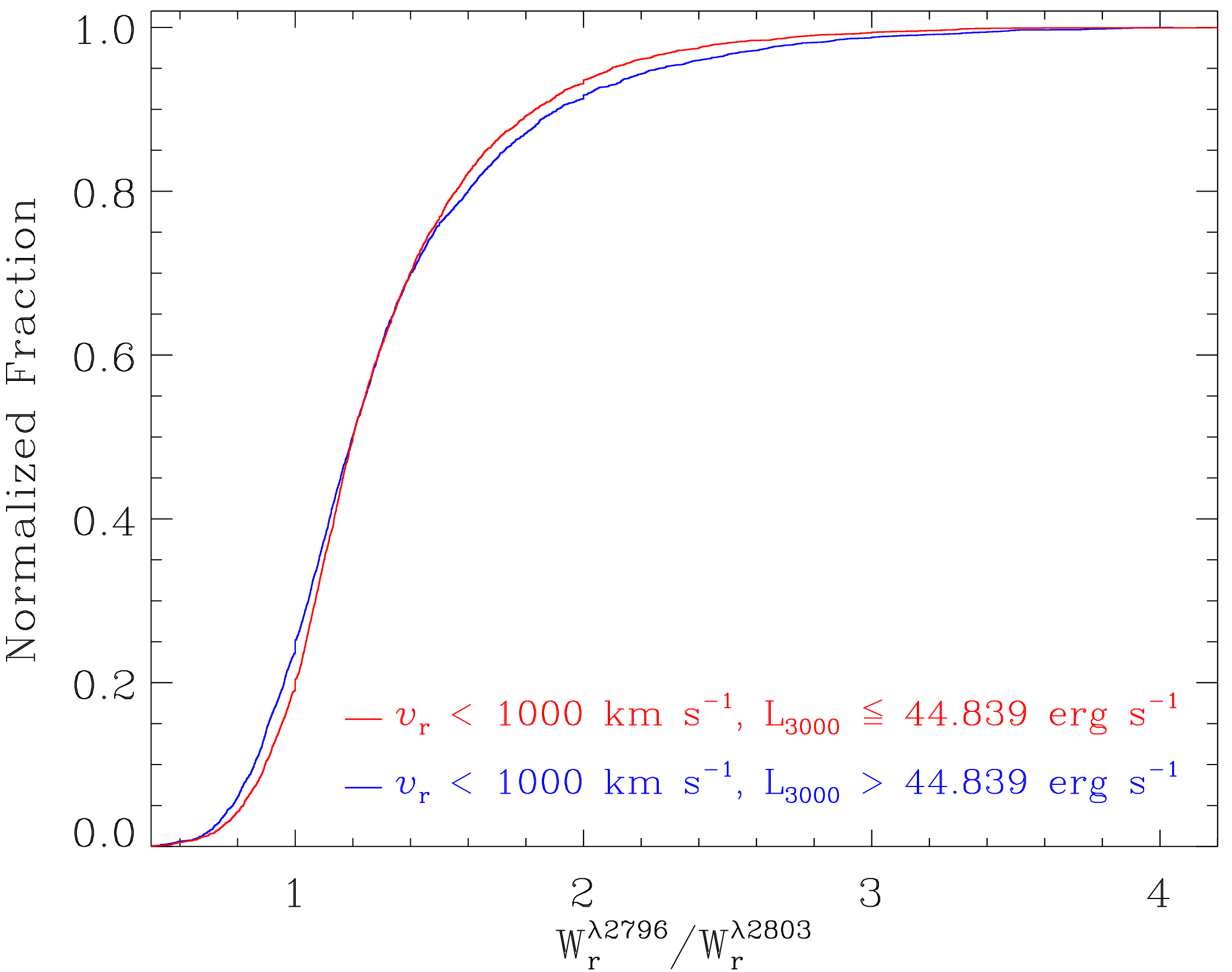}
\caption{Distributions of the absorption strength ratios of the \MgII\ doublets. Left panel: local data point densities with grayscale. The red solid-lines indicate the theoretical limits of completely saturated (DR$=W_{\rm r}^{\lambda2796}/W_{\rm r}^{\lambda2803}=1$) and unsaturated (DR$=W_{\rm r}^{\lambda2796}/W_{\rm r}^{\lambda2803}=2$) absorption, respectively. The blue solid-line represents the median values of $\rm DR$ as a function of $W_{\rm}^{\lambda2796}$. Middle and right panels are for the absorbers which are the same as those shown in the middle and right panels of Figure \ref{fig:distr_w}.}
\label{fig:distr_DR}
\end{figure*}

Figure \ref{fig:distr_DR} displays the ratio of the $W_r^{\lambda2796}/W_r^{\lambda2803}$ (DR), which reflects the saturated level of the \MgII\ absorption doublet. About 97.1\% of the \MgII\ absorbers are fallen into $1-\sigma_{\rm DR} \le DR \le 2+\sigma_{\rm DR}$, where 1 and 2 are the completely saturated an unsaturated absorption (Str{\"o}mgren 1948), respectively, and $\sigma_{\rm DR}$ is estimated via
\begin{equation}
\label{eq:DR}
\rm \sigma_{DR} \equiv \Delta DR = \frac{\Delta W_r^{\lambda2796}}{W_r^{\lambda2803}} + \frac{W_r^{\lambda2796}\Delta W_r^{\lambda2803}}{(W_r^{\lambda2803})^2}.
\end{equation}
The \MgII\ doublets are apt to the saturated absorptions with increasing $W_r^{\lambda2796}$ (see the blue line shown in Figure \ref{fig:distr_DR}). We do not find a noteworthy difference of $\rm DR$ between associated and intervening absorbers, and between faint and luminous quasars.

\section{Summary and future works}
\label{sect:summary}
Using the SDSS spectra of the $193~301$ unique quasars included in the DR7Q or DR12Q catalogs, this paper searches for narrow \MgIIab\ absorption doublets in the spectra data around \MgIIwave\ emission lines. We obtain $17~316$ \MgII\ absorption systems with velocity offset $\upsilon_r \le 10~000$ \kms\ from the quasars, which are imprinted in the spectra of $\rm 15~925$ unique quasars. Our main results and conclusions are as follows.
\begin{enumerate}
  \item Although there are a few associated \MgII\ absorbers which should have been accelerated to $\upsilon_r > 6000$ \kms, we find that the velocity $\upsilon_r < 6000$ \kms\ is a safe boundary to constrained a vast majority of associated \MgII\ absorbers, and about 6.8\% of quasars have at least one \MgII\ absorption doublets with $\upsilon_r < 6000$ \kms. While associated \MgII\ absorbers defined with $\upsilon_r < 6000$ \kms\ are significantly polluted by cosmologically intervening population, and about 66.7\% of the absorbers are likely belonged to the purely associated population, namely located within quasar outflow/wind, inflow and environment clumpy clouds. Therefore, we infer that only about 4.5\% ($66.7\% \times6.8\%$) of quasars present associated \MgII\ absorption systems with $W_{\rm r}^{\lambda2796}\ge0.2$ \AA.

  \item The fraction of associated \MgII\ absorption systems with high velocity outflows positively correlates with the average luminosities of their central quasars, which indicates a relationship between outflows and the quasar feedback power. In addition, the $\upsilon_r$ distribution of the outflow/wind \MgII\ absorbers is peaked at 1023 \kms, which is smaller than the corresponding value of the outflow/wind \CIV\ absorbers. The difference of peak velocities might be related to the ionization potentials and ionization mechanisms of the \MgII\ and \CIV\ absorbing gas.

   \item The profile of the redshift number density evolution of absorbers ($dn/dz$) limited by $\upsilon_r > -3000$ \kms\ or $\upsilon_r > 0$ \kms\ is different from that of absorbers constrained by $\upsilon_r > 2000$ \kms\ or higher values. While, absorbers limited by $\upsilon_r > 2000$ \kms\ and higher values exhibit similar profile of $dn/dz$. These phenomenons can be ascribed to absorption systems with $\upsilon_r > 2000$ \kms\ that are dominated by intervening absorbers, and to a significant contribution from associated absorbers if the absorption systems are limited by $\upsilon_r > -3000$ \kms\ or $\upsilon_r > 0$ \kms. In addition, we find that the $dn/dz$ is smaller when absorbers are constrained with larger $\upsilon_r$, which would be natural that the larger velocity limit the smaller fraction of associated absorbers is.

  \item In the hands of the absorption strength $W_{\rm r}^{\lambda2796}$ and strength ratio $W_r^{\lambda2796}/W_r^{\lambda2803}$ (DR), no significant difference is found between the associated and intervening absorbers, and between faint and luminous quasars.
\end{enumerate}

Up to this day, this paper provides us the largest sample of associated \MgII\ narrow absorption systems, which is very useful for us to investigate the quasar environment, feedback, an so on. In our future works, using the large associated \MgII\ absorber sample, we will investigate 1) the extinction of absorbers to quasar sightlines, 2) the relationships of the cluster characteristics between \MgII\ absorbers around quasars and galaxies around quasars, 3) incident rates of absorbers for different type quasars.

\acknowledgements We are grateful to the anonymous referee for careful comments that help to improve this manuscript. This work was supported by the National Natural Science Foundation of China (NO. 11363001; NO. 11763001), and the Guangxi Natural Science Foundation (2015GXNSFBA139004).

Funding for SDSS-III has been provided by the Alfred P. Sloan Foundation, the Participating Institutions, the National Science Foundation, and the U.S. Department of Energy Office of Science. The SDSS-III web site is http://www.sdss3.org/.

SDSS-III is managed by the Astrophysical Research Consortium for the Participating Institutions of the SDSS-III Collaboration including the University of Arizona, the Brazilian Participation Group, Brookhaven National Laboratory, Carnegie Mellon University, University of Florida, the French Participation Group, the German Participation Group, Harvard University, the Instituto de Astrofisica de Canarias, the Michigan State/Notre Dame/JINA Participation Group, Johns Hopkins University, Lawrence Berkeley National Laboratory, Max Planck Institute for Astrophysics, Max Planck Institute for Extraterrestrial Physics, New Mexico State University, New York University, Ohio State University, Pennsylvania State University, University of Portsmouth, Princeton University, the Spanish Participation Group, University of Tokyo, University of Utah, Vanderbilt University, University of Virginia, University of Washington, and Yale University.

\end{document}